\documentstyle[epsf,emulateapj]{article}

\slugcomment{Submitted to the Astrophysical Journal.}
\lefthead{HUANG \& SARAZIN}
\righthead{A4059 X-RAY EMISSION}

\begin{document}

\title{A High Resolution ROSAT X-Ray Study of Abell~4059}

\author{Zhenping Huang and Craig L. Sarazin}
\affil{Department of Astronomy, University of Virginia, \\
P.O. Box 3818, Charlottesville, VA 22903-0818; \\
zh2a@virginia.edu, cls7i@virginia.edu}

\begin{abstract}
We present the analysis of ROSAT HRI and PSPC X-ray data
on the compact cD type cluster Abell~4059. 
The central cD galaxy hosts the strong radio source 
PKS2354-35.
The central portion of the X-ray image shows 
two statistically significant X-ray emission minima and
an orthogonal X-ray bar.
As is the case in NGC~1275, 
the radio lobes in PKS2354-35 lie perpendicularly
to the X-ray bar and extend into the two X-ray minima.
One suggestion is that the radio lobe plasma has displaced 
the X-ray plasma and created the X-ray holes.
The radio plasma and the X-ray gas do have comparable 
pressures in the radio lobe area.
However, the available radio images do not show
radio emission which extends far enough to fill the two X-ray holes.
An alternative scenario for the anti-correlated radio and 
X-ray structure is an rotating disk in the cooling flow region 
of Abell~4059.

The ROSAT HRI surface brightness profile shows that there is a
cooling flow around the central cD galaxy with
a cooling rate of
$\dot{M} = 184^{+22}_{-25}$ $M_{\odot}$ yr$^{-1}$
within a radius of 156 kpc.
However, the ROSAT PSPC spectrum of the central regions 
does not require a cooling flow, and gives
an upper limit to the cooling rate of
$\dot{M} < 80 \, M_{\odot}$ yr$^{-1}$ (90\% confidence level).
The X-ray spectra of the central region
indicate very little intrinsic absorption in the cluster, 
with an upper limit of 
$\Delta N_H \le 4.0 \times 10^{19}$ cm$^{-2}$ for excess absorption
in front of the cluster emission at the center of the cooling flow.

On large scales, the cluster shows interesting
spatial alignment between 
the major axes of the X-ray emission, 
cD galaxy, and the cluster potential. 
This is consistent with the anisotropic merger model for the formation of
cD galaxies and the orientation of cD galaxy radio sources.
\end{abstract}


\section{INTRODUCTION} \label{sec:intro}

The Abell cluster A4059 is compact and 
dominated by a central cD galaxy ESO~349-G010.
The cluster is classified as richness~1
and is at a redshift of 0.0487 
(Abell, Corwin, \& Olowin 1989;
Green, Godwin, \& Peach 1990, hereafter GGP;
Schwartz et al.\ 1991).
The cluster is not well studied in optical.
Of the 284 galaxies which have been identified as candidates
for membership in the cluster,
only 11 galaxies have known redshifts
(Green et al.\ 1988, 1990).
The redshift of the cD galaxy appears to be 
consistent with the cluster mean velocity 
within the measurement errors (Green et al.\ 1990),
which may indicate that the cluster has achieved a 
somewhat relaxed state.

The central cD galaxy hosts the radio galaxy PKS2354-35,
with a total radio power of
$1.5\times10^{42}$ ergs s$^{-1}$
(Schwartz et al.\ 1991; Taylor et al.\ 1994).
A VLA A-array observation at 8.4 GHz (Taylor et al.\ 1994)
showed that the radio galaxy has a double-lobe structure
extending along the major axis of the host galaxy.
The alignment of the radio structure, cD major axis, 
and the overall cluster elongation is an intriguing 
feature, which may help us understand the formation 
and evolution of cD galaxies and the radio sources 
associated with these galaxies (West 1994). 

Observations with $HEAO~1$ and {\it EXOSAT} 
(Edge \& Stewart 1991;
Schwartz et al.\ 1991;
Edge, Stewart, \& Fabian 1992)
have established that the cluster is a bright X-ray source with
strong cooling flow around the central cD galaxy.
Recent {\it ASCA} observation also revealed a possible chemical
abundance gradient around the central cD galaxy
(Ohashi 1995).
A similar feature has been seen in a few other clusters,
such as the Centaurus, AWM7 (Ohashi 1995), 
and Virgo clusters (Matsumoto 1994).

Excess soft X-ray absorption has been seen in the
X-ray spectra of the central regions in a number of
cooling flow clusters of galaxies.
A4059 is near the southern galactic pole 
(galactic latitude of $-75.9^{\circ}$)
with a low galactic neutral hydrogen column density of
$1.45\times10^{20}$ cm$^{-3}$ 
(Stark et al.\ 1992). 
This makes A4059 a good case for studying
intrinsic X-ray absorptions.

In this paper, we will first analyze the
spatial properties of the global X-ray emission of the cluster 
(\S~\ref{sec:spatial}).
We then investigate the X-ray spectral properties of the gas,
such as the gas temperature distribution 
and the intrinsic absorption in the center of the cluster,
using ROSAT PSPC archival data (\S~\ref{sec:spectral}).
The detailed X-ray structure near the cD galaxy will be
compared to the radio emission from the cD galaxy in
\S~\ref{sec:central}.
Finally, we summarize the main results
(\S~\ref{sec:summary}).
Throughout the paper, we have assumed
a Hubble constant of 50 km s$^{-1}$ Mpc$^{-1}$ and $q_0$ = 0.5.

\section{OBSERVATIONS AND DATA REDUCTION} \label{sec:observ}

The cluster A4059 was observed with the ROSAT HRI 
(High Resolution Imager) on 6 October, 1993,
for 6263 seconds in four observation intervals (OBIs).
The observation was centered at the cD galaxy PKS2354-35.
The data reduction was done using the standard PROS package.
The data was first examined for periods of high background,
and about 140 seconds was removed. 
Vignetting and background corrections were applied.
A4059 has a relatively small angular size 
(see Fig.~\ref{fig:xray_galaxies}; 
Green et al.\ 1990).
The X-ray emission from the cluster is well confined within
the ROSAT HRI FOV ($\sim 40'$ in diameter).
Thus, we can determine the X-ray emission background from 
within the HRI field; 
the background region is chosen to be an annulus
centered at the cD galaxy PKS2354-35
with the inner and outer radii being 700$''$ to 900$''$, 
respectively.

\begin{figure*}[htb]
\vskip3.8truein
\includegraphics{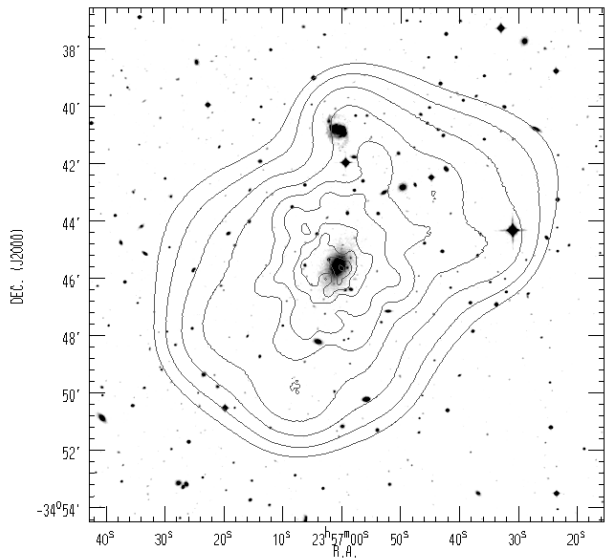}
\caption{
A contour map of the ROSAT HRI image of Abell~4059
superposed on an optical map from the Digitized Sky Survey.
The X-ray image was adaptively smoothed with a signal-to-noise
ratio of 5, and was corrected for background, vignetting, and exposure.
The X-ray contour levels are:
(0.000926, 0.00155, 0.00256, 0.00426, 0.00706, 0.0117, 0.0194, 0.0322,
 0.0534, 0.0886)
cnts arcmin$^{-2}$ s$^{-1}$.}
\label{fig:xray_dss_whole}
\end{figure*}

In order to provide spectral information to complement the
HRI spatial data, we retrieved the ROSAT PSPC observation of A4059.
The cluster was observed for 5514 seconds on 21-22 November, 1991
(observation WP800175; PI: R. Schwarz).
We first removed periods of high background from this data
(Master Veto rate $MV > 170$).
This resulted in a rejection of 107 seconds of exposure time.
The spectral fitting of the data was done using the XSPEC package. 

The pointing accuracy of the ROSAT HRI is important 
to our discussion of the central X-ray and radio structures 
in the cluster (\S~\ref{sec:central_radio}).
Unfortunately, there are no ideal X-ray reference sources
with accurate optical or radio positions in the HRI field
to calibrate the X-ray pointing accuracy.
The SB galaxy, ESO 349-G009 
(about $5'$ to the north of the cD galaxy)
may be associated with an X-ray source,
but the X-ray emission is extended,
which makes the position measurement of the X-ray source and the
correct optical correspondence uncertain.
There are several other galaxies 
to the southeast of the cD galaxy 
(Fig.~\ref{fig:xray_galaxies})
which may also have X-ray emission,
but these X-ray sources are too weak for their positions
to be determined accurately.
Thus, we rely on the absolute positions provided by the aspect
solution of the star camera on ROSAT.
Based on previous experience,
the HRI X-ray images presented in this paper could have
an absolute position error of up to about $10''$.
We discuss the relative alignment of the optical, radio, and
X-ray images in more detail in \S~\ref{sec:central_xray}.

\begin{figure*}[htb]
\vskip4.0truein
\includegraphics{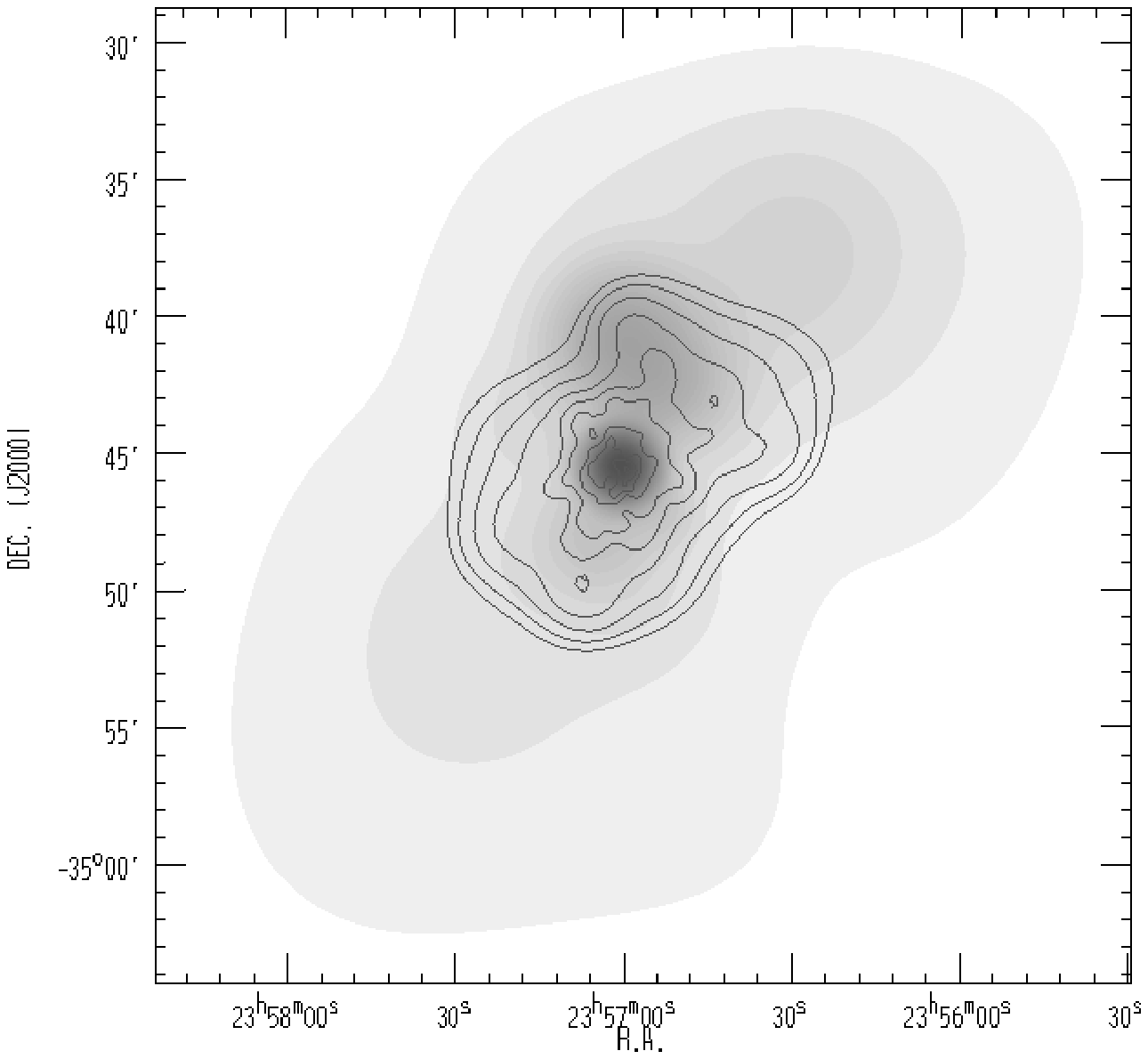}
\caption{
A contour map of the ROSAT HRI emission of Abell~4059 
superposed on a grey scale map of the smoothed galaxy $b$-band
optical surface brightness distribution.
The X-ray contour levels are the same as those in 
Figure~\protect\ref{fig:xray_dss_whole}.
The grey scale of the galaxy optical surface brightness
is logarithmically spaced, 
with the lowest and highest levels at
$6.85\times10^{-10}$ and $1.24\times10^{-7}$ 
ergs cm$^{-2}$ s$^{-1}$ arcmin$^{-2}$, respectively.}
\label{fig:xray_galaxies}
\end{figure*}

\section{GLOBAL X-RAY SPATIAL ANALYSIS} \label{sec:spatial}

\subsection{Global X-Ray Structure} \label{sec:global_struct}

As reported by {\it EXOSAT} (Schwartz et al.\ 1991) 
and {\it ASCA} (Ohashi 1995),
the X-ray emission from the cluster is concentrated near
the cD galaxy, PKS2354-35. 
Figure~\ref{fig:xray_dss_whole} shows the ROSAT HRI image 
of the entire cluster superposed on an optical image 
from the Digitized Sky Survey.
The X-ray image was corrected for background and vignetting,
and was adaptively smoothed with a signal-to-noise ratio of 5
per smoothing beam using our
adaptive kernel image smoothing routine 
(AKIS; Huang \& Sarazin 1996).

To better compare the global X-ray and optical properties of the cluster,
we also produced a smoothed galaxy optical surface brightness image
of the cluster, which is shown superposed on the contours 
of the X-ray emission in Figure~\ref{fig:xray_galaxies}.
In the smoothed optical surface brightness image, each galaxy was 
weighted by its $b$-band optical luminosity, and
the image was then adaptively smoothed with at least 25 average
galaxy luminosities per smoothing beam.
The galaxy positions and luminosities are from GGP.
We note that of the total of 284 galaxies listed as cluster members
by GGP, only 11 galaxies have the measured redshifts.
Therefore, some of the galaxies in the GGP sample 
may be nonmember galaxies seen in projection against the cluster.

Although the large scale X-ray image of A4059 is elongated with
NNW to SSE, there also is some evidence for a concentration in the
emission directly north of the central cD galaxy PKS2354-35.
In Figure~\ref{fig:xray_dss_whole}, this secondary peak appears
in the general direction of the SB galaxy ESO 349-G009.
With a blue magnitude of 14.6 (GGP), 
this is the second brightest galaxy in the cluster.
In the HRI image, there is a broad peak in this region.
The feature is stronger in the ROSAT PSPC image, but the
X-ray centroid lies about $30''$ to the south of the galaxy.
ESO~349-G009 was classified by Buta (1995) as a ring-type
galaxy, which may indicate an interaction with another galaxy and/or
strong star formation.
In fact, the galaxy is a powerful IRAS source (IRAS~23544-3457)
with a luminosity from 40 to 120 $\mu$m of
$7.5\times10^{10}$ L$_{\odot}$
(IRAS Point Source Catalog 1988).
The galaxy is also a radio source
with a flux density of $10.8\pm0.2$ mJy
at the 1.4 GHz band from the NVSS
(Condon et al.\ 1993),
which corresponds to $8.41 \times 10^{22}$ W Hz$^{-1}$
at the galaxy redshift ($z = 0.04207$).
The infrared and radio fluxes are consistent with the general correlation
found for starburst galaxies
(Condon 1992).
If ESO 349-G009 is the source of the extra X-ray emission to the north
of the cluster center, then its X-ray luminosity is approximately
$10^{42}$ ergs s$^{-1}$.
All of these properties are consistent with a strong starburst in
ESO 349-G009.
This galaxy may be similar to NGC~6045 in the Hercules cluster
(Huang \& Sarazin 1996).
This may indicate that there exist a class of spiral galaxies
in galaxy clusters
which are currently undergoing strong star formation at a rate
$\sim$10 times that of M~82.

Alternatively, Figure~\ref{fig:xray_galaxies} suggests that there is
a peak in the galaxy distribution to the north of the cluster center.
Because the grey scale image in this Figure gives the optical surface
brightness, part of this peak is due to ESO 349-G009.
However, there may be a subcluster in this region, which provides the
extra X-ray emission.

Figures~\ref{fig:xray_dss_whole} and \ref{fig:xray_galaxies} 
show that the global X-ray emission 
is elliptical and elongated along the major axis of the
galaxy luminosity distribution.
The major axis of the cD galaxy is also 
aligned in the same general direction. 

\begin{figure*}[htb]
\vskip3.5truein
\includegraphics{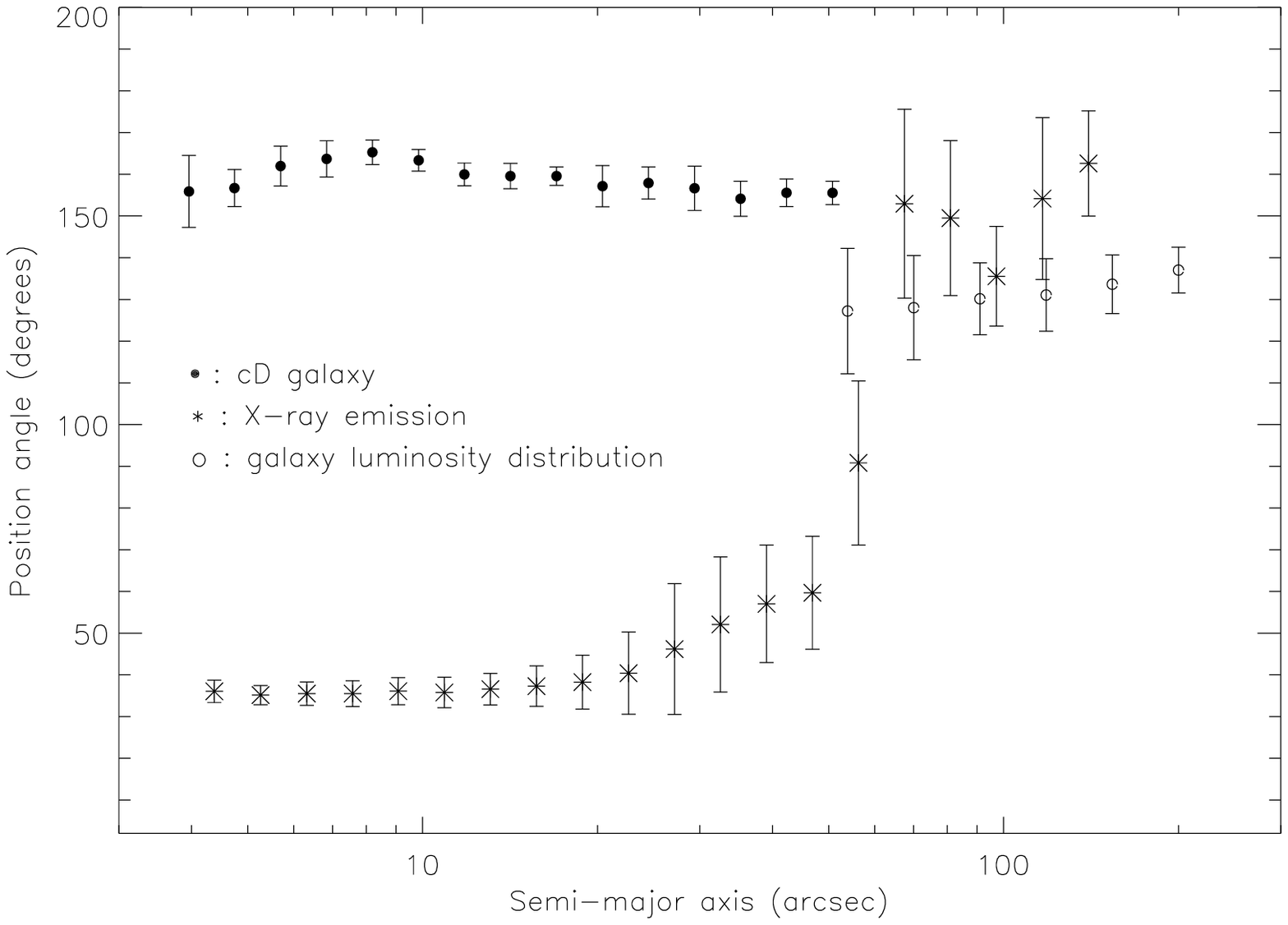}
\caption{
The position angles ($PA$) from elliptical fits to the X-ray surface
brightness, optical surface brightness of the cD galaxy PKS2354-35, 
and the cluster galaxy optical surface brightness as a function of the
semimajor axis.
$PA$ is measured counter-clockwise from the north.
The rapid change in the orientation of the X-ray emission at 40$''$
is due to the X-ray bar structure at smaller radii.}
\label{fig:ellipse_pa}

\vskip3.5truein
\includegraphics{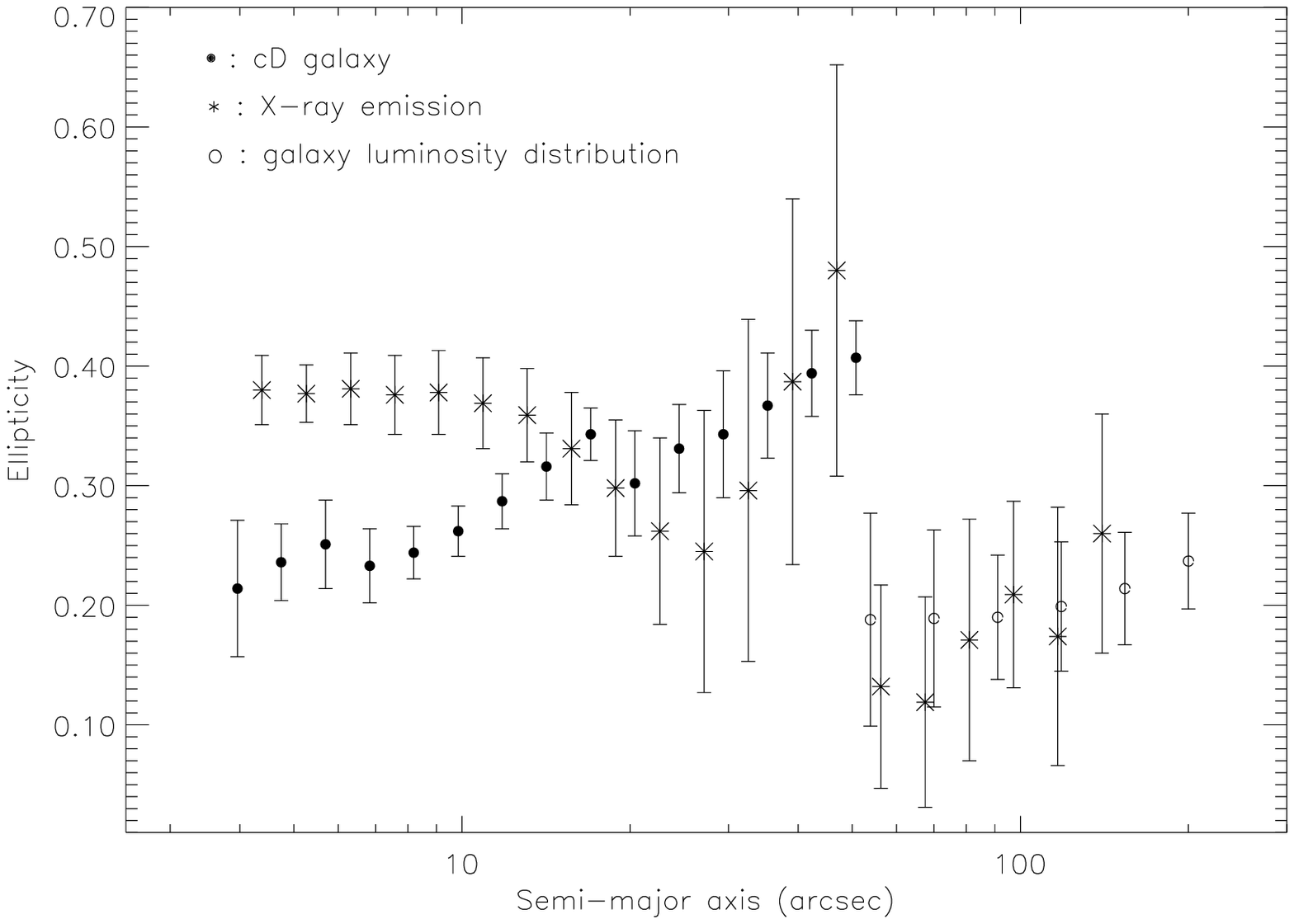}
\caption{
The ellipticity of the X-ray emission, optical surface brightness of
the cD galaxy PKS2354-35, and the cluster galaxy luminosity distribution
as a function of the semimajor axis for elliptical fits.
The ellipticity is defined as $1-b/a$, 
where $a$ and $b$ are the semimajor and semiminor axes of the ellipse, 
respectively.}
\label{fig:ellipse_e}
\end{figure*}

To quantify elongation of the cluster,
we have used the ellipse fitting routine in STSDAS
(Jedrzejewski 1987)
to fit elliptical isophotes to the X-ray emission, to the
cD galaxy optical surface brightness, and to the cluster
galaxy optical luminosity distribution.
The position angles ($PA$) and ellipticities are shown in 
Figures~\ref{fig:ellipse_pa}
and \ref{fig:ellipse_e}.
The $PA$ is measured counter-clock-wise from the north.
The optical image of the cD is from the Digitized Sky Survey.
Figure~\ref{fig:ellipse_pa} shows that 
the major axis of the cD galaxy is in 
$PA \approx 160^{\circ}$.
In the outer regions ($r \ga 50''$), the X-ray emission seems to have
the same orientation as the cD within the fitting errors.
A similar result was found from a wavelet analysis
of the ROSAT PSPC image (Slezak et al.\ 1994).
The outer X-ray orientation also agrees reasonably well with that
of the smooth galaxy optical surface brightness, and the X-ray
surface brightness and galaxy distribution give similar values for
the ellipticity, as well.
The galaxy distribution is at a slightly smaller position angle
than either the X-ray emission or cD galaxy (at smaller radii).
Our result for the orientation of the galaxy distribution
are consistent with those of GGP.

However, in the inner part ($r \la 50''$) of the X-ray image,
the elongation of the X-ray surface brightness is
is rotated by about 90 degrees with respect to the major axis of
the cD galaxy.
The X-ray image is also significantly more elliptical in the innermost
regions.
The sudden change in the $PA$ is
due to the X-ray bar structure in the central area.
We will discuss this feature in detail in \S~\ref{sec:central}.

\subsection{X-ray Surface Brightness and Gas Density Profiles}
\label{sec:spatial_profile}

To derive the X-ray surface brightness profile,
the X-ray emission was accumulated in circular annuli.
Tests have shown that the results for circular annuli are
consistent with the results for elliptical annuli 
for moderate values of the ellipticity
(White et al.\ 1994).
The X-ray counts in these bins were corrected for vignetting, 
exposure, and background.
The count rate was then converted into a physical flux in
the ROSAT X-ray band $0.1-2.4$ keV, 
assuming a Raymond-Smith thermal spectrum
with a temperature of 3.43 keV
(\S~\ref{sec:spectral_temp}),
the total (galactic and intrinsic) absorption column density
$N_H = 1.40\times10^{20}$ cm$^{-2}$
(\S~\ref{sec:spectral_cool}),
and the abundances from the {\it ASCA} spectrum
(Ohashi 1995).

\begin{figure*}[htb]
\vskip4truein
\includegraphics{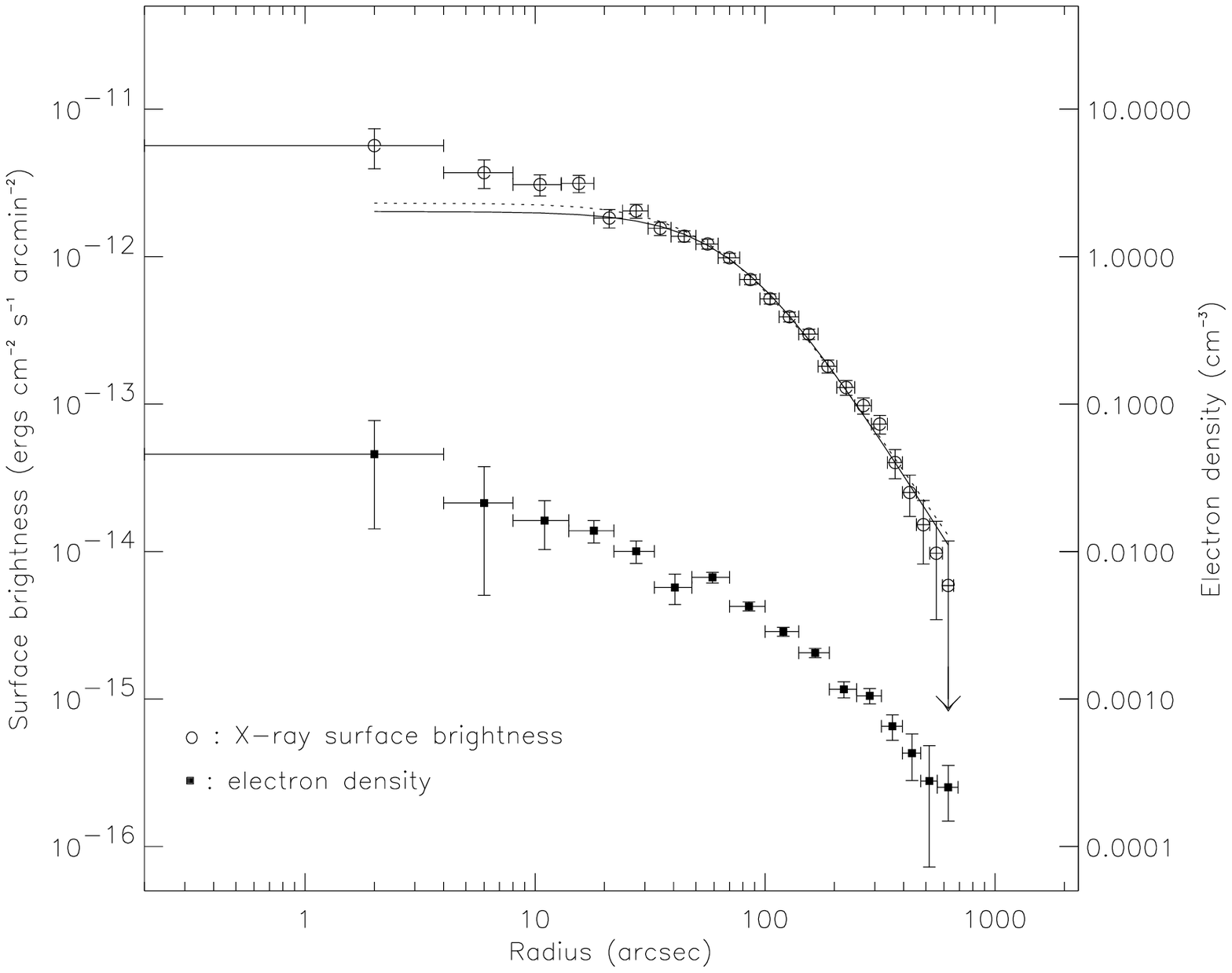}
\caption{
The ROSAT HRI X-ray surface brightness and de-projected electron
density for A4059. 
The upper data (open circles) give the X-ray surface brightness in
the photon energy band 0.1$-$2.4 keV 
(vertical scale on the left).
The solid curve is the best-fit $\beta$-model 
for the surface brightness with all 23 data points,
while the dotted line is the fit with the first 4 data points removed.
Both fits show that these is an excess X-ray emission 
in the center of the cluster.
The lower data points (filled squares) 
are the de-projected electron density distribution
(scale on the right).
The error bars show the 1-$\sigma$ errors for the surface brightness 
or the electron density. 
For the X-ray surface brightness,
the errors are derived directly from Poisson counting statistics;
the errors for the electron density are from 100 Monte Carlo
simulations.}
\label{fig:xray_surf_whole}
\end{figure*}

The azimuthally averaged X-ray surface brightness profile
is plotted in Figure~\ref{fig:xray_surf_whole} 
(upper curve).
The errors for each data point follow
directly from Poisson counting statistics including the
background subtraction.
 From the X-ray surface brightness,
one can derive the electron density distribution
by de-projection (Fabian et al.\ 1981).
Spherical symmetry was assumed,
and the same thermal spectrum was adopted as used in the conversion to
the physical flux.
The resulting electron density profile is shown as the lower set of
filled squares in Figure~\ref{fig:xray_surf_whole}.
The errors on the electron density are 68\% confidence ranges
derived from Monte Carlo simulations of 100 statistical realizations
of the data. 

In Figure~\ref{fig:xray_surf_whole}, 
the X-ray surface brightness is strongly peaked
toward the center of the cluster.
We tried to fit the data using the
conventional isothermal $\beta$-model,
$I_X (r) = I_0 \ [1 + ( r / r_X )^2]^{-3\beta + 1/2}$.
Here, $I_0$ is the central surface brightness, 
$r$ is the projected radius, 
and $r_X$ is the core radius of the X-ray emission.
The model was convolved with the Point Response Function (PRF) 
of the ROSAT HRI.
The best $\chi^2$ fit for all 23 data points
gave $\beta = 0.55^{+0.03}_{-0.03}$ 
and $r_X = 66^{+10}_{-9}$ kpc 
(uncertainties are 90\% confidence limits for
one interesting parameter).
The value of $\chi^2$ was 27.8 for 20 degrees of freedom (d.o.f.),
which is not an acceptable fit.
The model surface brightness (solid line) is plotted in
Figure~\ref{fig:xray_surf_whole}.
Clearly, there is a central excess of X-ray emission 
compared to the model for the first four data points.
To test the significance of this excess,
we removed the four innermost data points
from the fit.
The resultant $\chi^2$ value was significantly reduced 
by 17.5 to 10.3 for 16 d.o.f.,
which is a dramatic reduction for the removal of four data points.
The improved fit, which gives
$\beta = 0.58^{+0.03}_{-0.04}$ and 
$r_X = 77^{+11}_{-12}$ kpc (90\% confidence level),
is plotted as the dotted line in Figure~\ref{fig:xray_surf_whole}.
Ohashi (1995) fitted a $\beta$-model 
to his {\it ASCA} data on the X-ray
surface brightness, and found larger values of
$\beta$ (0.65) and $r_X$ (100$''$).
The difference may be due in part to the larger PRF
or harder spectral band of {\it ASCA}.

\subsection{Cooling Flow} \label{sec:global_cflow}

The excess central X-ray surface brightness suggests that A4059
has a cooling flow, as was previously found by 
Schwartz et al.\ (1991) and
Edge et al.\ (1992).
We find that the average electron density 
within the central $4''$ is
$n_e \approx 0.046\pm0.035$ cm$^{-3}$,
which implies that the gas cooling time for this region is about
$3.1^{+17.2}_{-1.8} \times 10^8$ years 
at a temperature of $1.93^{+0.76}_{-0.33}$ keV 
(90\% confidence level).

 From the electron density profile, we find that the
cooling radius, defined as the point where the cooling time
equals $10^{10}$ yr, is 
$r_c = 120^{+14}_{-9}$ arcsec = $156^{+18}_{-12}$ kpc 
(90\% confidence level).
The total cooling rate within this radius is
$\dot{M} = 184^{+22}_{-25}$ $M_{\odot}$ yr$^{-1}$
(90\% confidence level).
This cooling rate is intermediate between the value of
$\dot{M} = 124^{+53}_{-44} \ M_{\odot}$ yr$^{-1}$
from Edge et al.\ (1992) and the estimate of
320 $M_{\odot}$ yr$^{-1}$ by Schwartz et al.\ (1991),
both based on {\it EXOSAT} data.

\section{SPECTRAL ANALYSIS OF THE X-RAY EMISSION} \label{sec:spectral}

\subsection{ Gas Temperature Distribution } 
\label{sec:spectral_temp}

We used the ROSAT PSPC observation of A4059 from the archive
(see \S~\ref{sec:observ}) to determine the (projected) X-ray
spectra in circular annuli centered on the cD galaxy PKS2354-35.
The results of the spectral fitting are summarized in 
Table~\ref{tab:spectral}.
Because of the limited spatial resolution of the ROSAT PSPC,
the width of each annular bin was set at least 1 arcmin.
The radii are given in the first row of 
Table~\ref{tab:spectral}.
For outer bins, the width was increased so that each bin had
enough ($>$ 1000) net photons in the spectral fitting.
The spectra were corrected for particle and X-ray background;
the average Master Veto rate was $MV = 79.88$.
The number of net photons in the spectrum after background subtraction
are listed in the second row of Table~\ref{tab:spectral}.
The spectra were limited to the energy channels
from 0.2 to 2.04 keV.

\begin{table*}[ht]
\caption[Spectral Fits of the ROSAT PSPC Archive of A4059]{}
\label{tab:spectral}
\begin{center}
\begin{tabular}{lcccccc}
\multicolumn{7}{c}{\sc ROSAT PSPC Spectral Fits} \cr
\tableline
\tableline
Radius ($''$)     &  0 -- 60 & 60 -- 120 & 120 -- 180 & 180 -- 260 
                  &   260 -- 410 & 410 -- 800                        \cr
Net Photons       & 1914 & 2283 & 1628 & 1460 & 1707 & 1531          \cr
Abundance         & 0.63 & 0.56 & 0.53 & 0.49 & 0.43 & 0.29          \cr
Temperature (keV) & 1.93$^{+0.76}_{-0.33}$ & 2.93$^{+1.32}_{-0.73}$  
                  & 4.64$^{+3.16}_{-1.64}$ & 2.76$^{+1.75}_{-0.89}$ 
                  & $> 3.03$ & $> 2.60$                              \cr
$\chi^2$/d.o.f.   & 51.5/67 & 80.9/77 & 54.4/60 & 55.5/54 & 62.8/62 
                  & 65.8/56                                          \cr
\tableline
\end{tabular}
\end{center}
\end{table*}

A Raymond-Smith thermal spectrum was assumed
with a Galactic neutral hydrogen column density of
$N_H = 1.28^{+0.18}_{-0.20}\times10^{20}$ cm$^{-2}$
(90\% confidence level; see \S~\ref{sec:spectral_cool}).
Because the ROSAT PSPC data did not provide strong constraints
on the abundances in the gas, we have fixed the 
abundances at different radii to values interpolated from the
{\it ASCA} measurements (Ohashi 1995).
These abundance values are listed in the third row of
Table~\ref{tab:spectral}.
The best-fit temperature values are given in row 4 of the Table;
the upper and lower limits of the gas temperatures 
are at the 90\% confidence level.
Because of the large fitting errors,
only lower limits are given for temperatures in bin 5 and 6.
The last row in the Table gives the total $\chi^2$ value 
for the fit and the number of degrees of freedom 
for each annular bin.

\begin{figure*}[htb]
\vskip3.5truein
\includegraphics{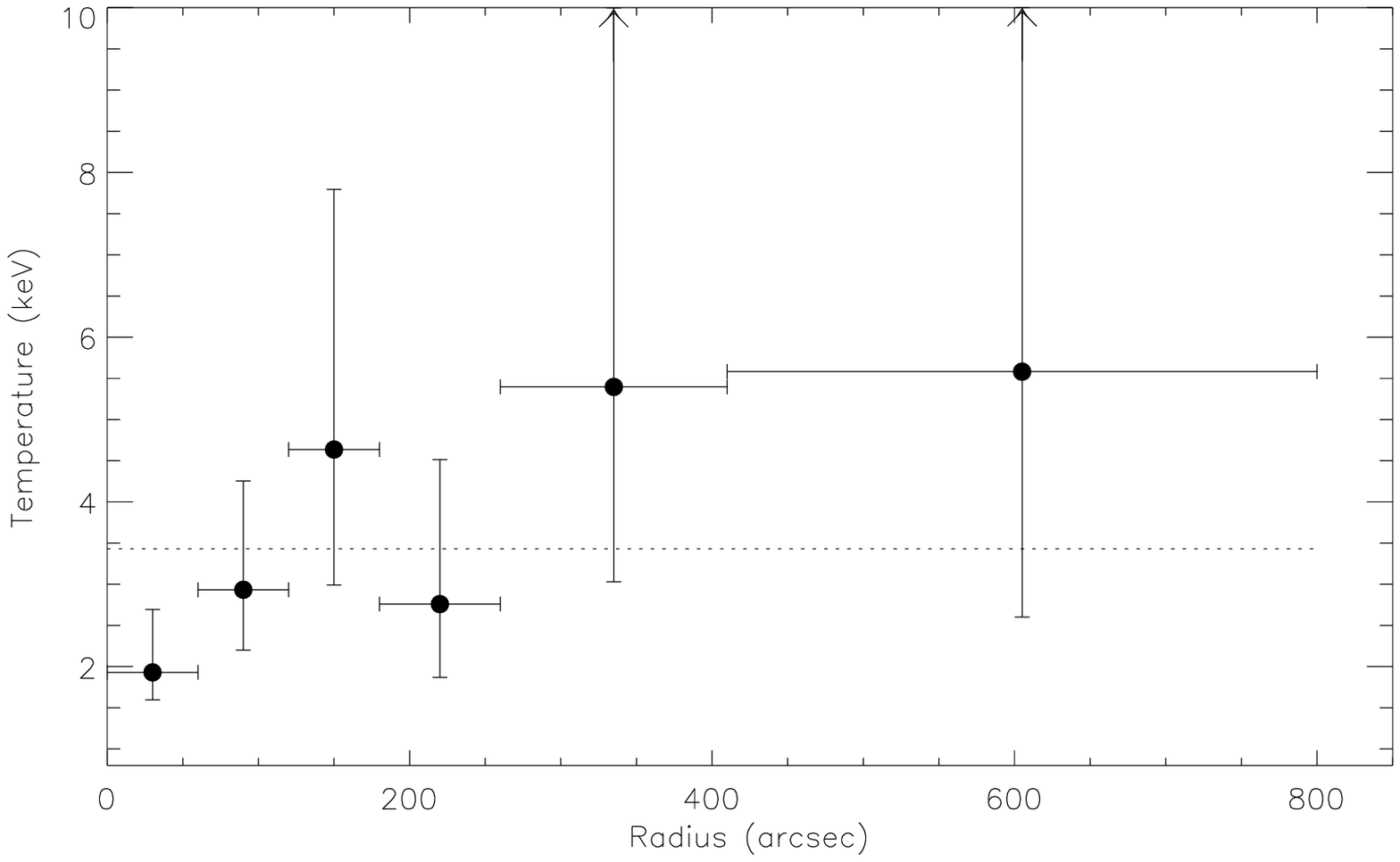}
\caption{
The azimuthally averaged temperature distribution for the X-ray gas
in A4059.
The temperatures are derived from the ROSAT PSPC data
assuming Raymond-Smith thermal spectrum for the X-ray gas
(Table~\protect\ref{tab:spectral}).
The dotted line is the best single temperature fit
for the entire cluster (3.43 keV).
The errors give the 90\% confidence intervals for the
temperatures.}
\label{fig:pspc_temperature}
\end{figure*}

The temperature distribution of the gas is plotted
in Figure~\ref{fig:pspc_temperature}.
Also plotted in the Figure is the result (dotted line)
of a single temperature fit of the entire cluster
within a radius of 800$''$.
This single temperature fit gives a
total $\chi^2$ of 175.3 for 162 degrees of freedom.
It gives the gas temperature and metal abundance of
$3.43^{+1.06}_{-0.73}$ keV and
$0.28^{+0.32}_{-0.24}$ solar (90\% confidence), respectively. 
It also gives the best fit value for the absorption column of
$N_H = 1.40^{+0.20}_{-0.20}\times10^{20}$ cm$^{-2}$,
which is consistent with the value from Stark et al.\ (1992).
This result is in agreement with
the results from other instruments:
$3.7^{+0.7}_{-0.7}$ keV (1~$\sigma$ confidence;
$EXOSAT$; Schwartz et al.\ 1991;  Edge \& Stewart 1991),
$3.5^{+0.3}_{-0.4}$ keV 
(1~$\sigma$ confidence; $Einstein$; David et al.\ 1993),
and 3.9 keV ({\it ASCA}; Ohashi 1995).

Figure~\ref{fig:pspc_temperature} shows that,
except the central annular bin,
the single temperature fit is a fairly good representation of the 
gas temperature distribution in the cluster.
The central bin seems to suggest a gas temperature drop in the
center of the cluster.
To test this hypothesis, we have divided the entire cluster
into two sets of spectral data: 
the central bin and the rest of the cluster.
We then fit the two spectra simultaneously, either
requiring that the temperature of the central bin equal that of
the rest of the cluster, or allowing the central bin to have a
different temperature.
Allowing the central bin to have a separate temperature
resulted in a reduction of the total $\chi^2$ value by 7.9
(from 218.3 to 210.4)
with the decrease of 1 degree of freedom 
(from 222 to 221).
 From the $F$-test, the hypothesis that the central temperature
is the same as that of the rest of the cluster can be rejected at the
95\% probability level.

\subsection{Gas Cooling and Intrinsic X-ray Absorption} 
\label{sec:spectral_cool}

The spatial analysis of the ROSAT HRI X-ray surface brightness
requires that there be a cooling flow in the central region of A4059
(\S~\ref{sec:spatial}).
The reduction in the central temperature is also consistent with this
(see \S~\ref{sec:spectral_temp}, Figure~\ref{fig:pspc_temperature}, 
and the preliminary ASCA result by Ohashi [1995]).
However, the ROSAT PSPC spectra do not require any significant
cooling flow component.
For example, within $r < 150''$ 
(slightly larger than the cooling radius; \S~\ref{sec:spatial}),
a single temperature Raymond-Smith thermal model 
gives a reasonably good fit to the spectra:
total $\chi^2$ of 145.6 for 135 degrees of freedom.
The addition of the cooling flow component
to the above model does not improve the fit.
The maximum cooling rate within this radius is 
$\dot{M} < 29 \, M_{\odot}$ yr$^{-1}$ at the 90\% confidence level.
For the PSPC spectra of the entire cluster 
($ r < 800''$), 
the addition of a cooling flow component does not
improve the corresponding single temperature model fit either.
At 90\% confidence level, the cooling rate 
is $\dot{M} < 80$ $M_{\odot}$ yr$^{-1}$
with a fitted absorption column density of
$N_H = 1.39^{+0.11}_{-0.13}\times10^{20}$ cm$^{-2}$.
This strong upper limit on the cooling rate is clearly 
inconsistent with the result from the spatial data analysis. 

One explanation for this inconsistency between the
X-ray surface brightness profile and the X-ray spectrum of 
the central region of the cluster 
would be if the cooling flow had only begun recently.
In \S~\ref{sec:global_cflow}, 
we determined the gas cooling time at the center of the cluster. 
Because of the large errors in the electron density,
the cooling time of the central gas could be as large as 
$2.0 \times 10^9$ years.
This is comparable to the gas relaxation time following
a subcluster merger (Roettiger, Burns, \& Loken 1993).
Thus, if this cluster recently underwent a merger which disrupted any
pre-existing cooling flow, it is possible that cooling of the central gas
might have only begun within the last $\la$2 Gyr.
Then, very little gas may have actually cooled down to low 
temperatures ($k T \la 10^7$ K).
On the other hand, A4059 now appears to be a quite relaxed cluster.
The X-ray image is quite regular on large scales
(Fig.~\ref{fig:xray_dss_whole}; Buote \& Tsai 1996),
although Slezak et al.\ (1994) has suggested that there is some evidence
for substructure based on a wavelet analysis of the ROSAT PSPC image.

The spectral fitting also shows
there is very little (if any)
intrinsic, excess soft X-ray absorption toward the center of the cluster.
If we adopt the Galactic neutral hydrogen column density of 
$N_H = 1.45\times10^{20}$ cm$^{-2}$ 
(Stark et al.\ 1992), 
the limit on the intrinsic absorption in the center of the cluster 
($r \le 60''$)
is 
$\Delta N_H < 0.14\times10^{20}$ cm$^{-2}$ 
at the 90\% confidence level.
Alternatively, we can use the spectrum of the outer cluster 
($r = 125'' - 800''$)
to fix the Galactic absorption, 
and determine the excess absorption in the center. 
The best-fit model gives central excess absorption
$\Delta N_H = 0.088\times10^{20}$ cm$^{-2}$
with a total $\chi^2$/d.o.f. = 192.1/207.
At 90\% confidence level, the central excess intrinsic absorption 
has only an upper limit of
$\Delta N_H < 0.40\times10^{20}$ cm$^{-2}$.
This upper limit for A4059 is rather low when compared to 
X-ray spectral fits of some other cooling flows clusters
(White et al.\ 1991; Allen et al.\ 1995).
However, this limit only applies to excess absorption which is
in front of the cluster emission.

\section{CENTRAL STRUCTURE OF THE X-RAY AND RADIO EMISSION} 
\label{sec:central}

\subsection{Central X-ray Structure} \label{sec:central_xray}

\begin{figure*}[htb]
\vskip4truein
\includegraphics{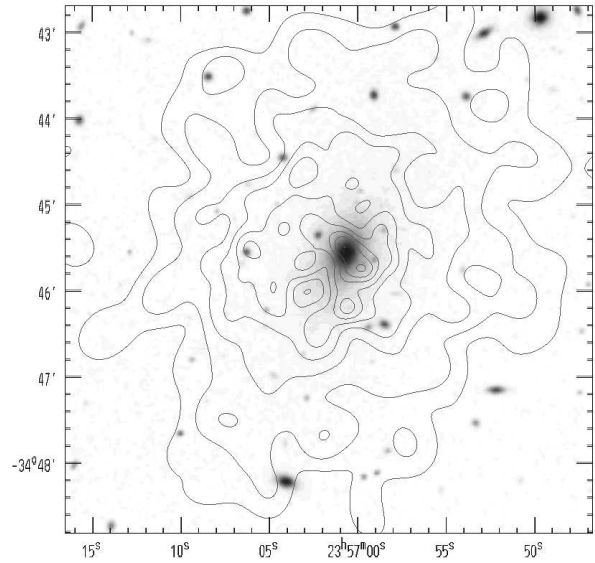}
\caption{
The central structure of the ROSAT HRI X-ray emission in A4059.
The X-ray contours are superposed on the optical image from
the Digitized Sky Survey.
The X-ray image was adaptively smoothed 
to a signal-to-noise ratio of 5 per smoothing beam.
The X-ray contour levels are:
(0.0108, 0.0148, 0.0202, 0.0275, 0.0374, 0.0510, 0.0694, 0.0946, 0.129)
cnts arcmin$^{-2}$ s$^{-1}$.
Two features about 1/2 arcmin to the southeast and northwest 
of the center of the radio galaxy are actually holes rather
than peaks (see Figure~\protect\ref{fig:xray_central_radio} below).}
\label{fig:xray_central_opt}
\end{figure*}

In the center of the cluster, the X-ray emission is 
strongly peaked towards the cD galaxy PKS2354-35.
Figure~\ref{fig:xray_central_opt} shows the contours of the
X-ray emission within the central 6$'$ 
superposed on the corresponding optical field 
from the Digitized Sky Survey.
The X-ray emission was adaptively smoothed 
with a signal-to-noise ratio of 5 per smoothing beam.
As discussed in \S~\ref{sec:global_struct},
the X-ray emission is elongated in the same direction 
as the cD on the scales of $r \ga 50''$.
However, on smaller scales ($r \la 50''$), the X-ray
emission is concentrated to an elongated X-ray bar,
which is nearly perpendicularly to the major axis of the cD.
The centermost part of the X-ray image is shown in grey scale in 
Figure~\ref{fig:xray_central_radio},
superposed on the 4.8 GHz radio contour map 
(Taylor et al.\ 1994).
The radio source will be discussed in the next section 
\S~\ref{sec:central_radio}.
As discussed in \S~\ref{sec:observ},
we have adopted the absolute positions from the aspect solution
for the HRI X-ray data.
The agreement of the radio and optical nucleus with a central position
along the X-ray bar suggests that the X-ray positions are not completely
wrong.
However, the brightest part of the bar extends for about 15$''$, so
this is not a very strong constraint on the accuracy of the X-ray
positions.

Figure~\ref{fig:xray_central_radio} also shows 
two X-ray emission minima (``X-ray holes'')
above and below the X-ray bar.
Are the two X-ray holes statistically significant?
First, we compare the X-ray surface brightness 
in the two X-ray holes with the average surface brightness 
at the same radius.
The two X-ray holes are defined as two elliptical regions
with semimajor and semiminor axes of $13''$ and $10''$, 
centered at 
R.A.\ = $23^{\rm h}57^{\rm m}02{\fs}8$,
Dec.\ = $-34^{\circ}46^{\prime}02{\farcs}5$,
and
R.A.\ = $23^{\rm h}56^{\rm m}59{\fs}7$,
Dec.\ = $-34^{\circ}45^{\prime}02{\farcs}0$
A total of 15 and 12 photons were detected in the
northern and southern holes, respectively.
Based on the counts in an annulus between $24''$ and $50''$,
we expect $29.7 \pm 1.4$ counts in each X-ray hole.
Thus, the Poisson probabilities that the number of photons 
in each X-ray hole
could be equal or less than 15 and 12 are about
0.23\% and 0.019\%, respectively.

Alternatively, we can verify the statistical significance 
level of the two X-ray holes by 
performing the Poisson simulations of the central X-ray image 
based on the ellipse fitting model (\S~\ref{sec:spatial}).
Out of 1000 statistical realizations of the image,
we found three instances in that the simulated X-ray count 
in the northern X-ray hole was less than or equal to 15, and
no cases where the simulated X-ray count in either hole was
less than or equal to 12.
These tests suggest that the two X-ray holes are real, 
and are unlikely to be the results of statistical fluctuations.

\subsection{Radio Lobes and Two X-ray Holes} \label{sec:central_radio}

\begin{figure*}[htb]
\vskip3.3truein
\includegraphics{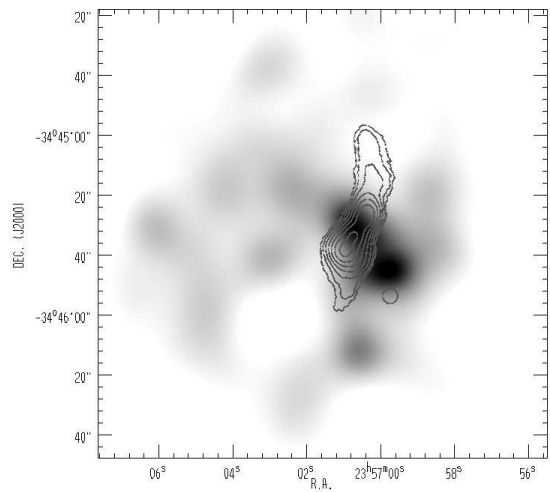}
\caption{
Contours of the VLA 4.8 GHz radio image (Taylor et al.\ 1994)
are superposed on a grey scale version of the ROSAT HRI X-ray map. 
The X-ray grey scale ranges from
0.0133 to 0.136 cnts arcmin$^{-2}$ s$^{-1}$.
The intensities near the centers of the northern and
southern X-ray holes are 
0.0265 and 0.0199 cnts arcmin$^{-2}$ s$^{-1}$,
respectively.}
\label{fig:xray_central_radio}
\end{figure*}

The central cD galaxy PKS2354-35 is a strong radio source
with two relatively bright radio lobes 
(the radio jets  are not pronounced).
A 4.8 GHz B-array VLA radio contour map is shown in 
Figure~\ref{fig:xray_central_radio} 
(Taylor et al.\ 1994).
The linear size of the radio source is also measured
from a VLA 1.4 GHz map from the NRAO VLA Sky Survey (NVSS),
which shows a slightly larger linear size 
(after deconvolution with the radio beam)
of about $63'' \times 22''$ (FWHM).

It is interesting that the two radio lobes lie 
perpendicular to the central X-ray bar 
and extend towards the areas of the two X-ray minima. 
The radio and X-ray structure in A4059
is similar to that of NGC~1275 in the Perseus cluster 
(B\"{o}hringer et al.\ 1993),  
it was argued that the two X-ray minima are due
to the displacement of X-ray emitting gas by the radio plasma. 
However, in A4059 the radio emission at 1.4 GHz and 4.8 GHz
does not extend into the central areas of the two X-ray holes,
although it is possible that the radio source 
extends further at lower frequencies. 

In the following, we compare the nonthermal pressure 
in the radio lobes with thermal pressure of the 
X-ray emitting gas in the region around the radio lobes.
Because we are most interested in the interaction of the radio
plasma and the X-ray emitting gas on the scale of the X-ray
structures seen in Figure~\ref{fig:xray_central_radio},
we will estimate the radio pressure on the largest observed scales
of the radio.
The radio lobes end at about 35$''$ from the center, 
where the average electron density is about 
$8.6\pm2.5 \times 10^{-3}$ cm$^{-3}$ 
(90\% confidence level).
The X-ray gas thermal pressure is then
$5.2^{+3.5}_{-2.3}\times 10^{-11}$ ergs cm$^{-3}$ at this radius,
assuming a gas temperature of $1.93^{+0.76}_{-0.33}$keV
(Table~\ref{tab:spectral}).
The largest scale of the radio emission from the
1.4 GHz NVSS radio image 
(Condon et al.\ 1993)
is about $63'' \times 22''$ (FWHM).
PKS2354-35 is a 1.27 Jy source at 1.4 GHz,
and has a spectral index of $-1.43$
(Taylor et al.\ 1994).
Assuming that this extended radio flux 
comes from a uniform prolate cylindrical region of size
$63'' \times 22''$,
that the ratio of the energy of ions to electrons is unity,
that the volume filling factor is unity,
that the magnetic field is perpendicular to the line of sight,
and that the lower and upper cutoff of the radio frequencies 
are 10 MHz and 100 GHz, respectively,
we find a minimum energy magnetic field of 17 $\mu$G, 
and a radio pressure
of $9.0\times10^{-12}$ ergs cm$^{-3}$
(e.g., Miley 1980),
which is about a factor of 5.8 smaller than the
gas thermal pressure. 
But considering the uncertainties in the estimate of the
minimum-energy radio pressure
(e.g., if the energy ratio of ions to electron is 100,
the radio pressure becomes $8.4 \times10^{-11}$ ergs cm$^{-3}$),
the radio pressure may be comparable to 
that of the thermal X-ray gas.

Closer to the center, nonthermal radio pressures 
are somewhat higher in the two radio lobes; 
the pressures are
$3.1 \times 10^{-11}$ ergs cm$^{-3}$ and
$5.6 \times 10^{-11}$ ergs cm$^{-3}$
for the northern and southern radio lobes, respectively
(Taylor et al.\ 1994).
These pressures are, however, measured at a distance of
$6.8'' - 8.5''$ from the radio core.
On this scale, the X-ray thermal pressure is about
$1.1^{+1.3}_{-1.0} \times 10^{-10}$ ergs cm$^{-3}$.
So, the thermal and nonthermal pressures are again comparable
in the innermost regions,
and it is possible that the radio lobes and the X-ray gas 
are in pressure equilibrium.
This would be expected if the radio lobes have displaced, expanded
against, and been confined by the X-ray gas.
This is consistent with the interpretation of the X-ray holes as regions
where the radio source has displaced the X-ray gas.

On the other hand, it is odd that the radio emission in 
the available images does not appear to fill the observed 
X-ray holes.
One possibility is that the relativistic electrons in the X-ray 
hole regions have undergone severe synchrotron losses,
or that the X-ray holes are residuals of
a previous epoch of radio activity.
In a 17 $\mu$G magnetic field, synchrotron losses dominate,
and the lifetime of electrons emitting at a frequency of 1.4 GHz,
the lowest frequency at which the source was observed and detected
with on the largest linear size, is $1.2 \times 10^7$ years.
We can also estimate the age of the radio source from
the variation in the radio spectral indices
(Myers \& Spangler 1985).
Taylor et al.\ (1994) found that the spectral index of the radio 
source changed from $-1.1$ at the radio core to $-2.5$ at
the faint, diffuse component to the radio lobes,
which suggests this faint, diffuse radio component
has been around for about $2.6 \times 10^7$ years.
Thus, we are not surprised that much of the
radio emission at 1.4 GHz or higher 
has faded beyond the region of the observed radio lobes.
An upper limit to the age of the X-ray holes, assuming they are supported
by nonthermal pressure, is the timescale for their disruption by
Rayleigh-Taylor instabilities and/or by buoyant forces.
Roughly, either of these processes should occur on a timescale which is
$\sim$10 times the sound crossing time of the holes in the
surrounding X-ray gas.
This maximum age is about $10^9$ years, which is
much longer than the lifetime of electrons emitting
at the observed radio frequencies.
Thus, it is possible that
the radio emission from much of the region has faded,
but that the lobes are still supported by the pressure of cosmic ray ion,
low energy cosmic ray electron, and the magnetic field.
In general, in a radio source with a steep spectrum like 
PKS2354-35 (Taylor et al.\ 1994),
most of the nonthermal electron pressure is due to electrons 
at low energies, which might emit at very 
low radio frequencies.

\subsection{Alignment of the X-ray, Optical, and Radio Emission}
\label{sec:central_alignment}

As described in \S~\ref{sec:spatial}, 
the X-ray isophotes for $r \ga 50''$ are aligned with the
direction of elongation of the optical isophotes of the cD galaxy
PKS2354-35, and the cluster galaxy isopleths.
In addition, the strong radio lobes from the cD galaxy 
is also extended along the major axis of the cD potential.
This is consistent with the anisotropic merger model 
for the formation cD galaxies and the orientation of their sources
as proposed by West (1994).
West noted that the alignments in A4059 continue to larger scales,
where the two nearest Abell clusters also lie approximately 
along the same axis.
In his model, cD galaxies were formed by anisotropic mergers occurring
preferentially along an axis determined by large scale structure.
Gaseous accretion associated with these mergers determined the
axis of rotation of a central black hole, and this is the axis along
which radio jets propagate.
Evidence for the alignment of the X-ray emission, central cD galaxy,
subclusters, host clusters, and surrounding superclusters
have been proposed recently in many other clusters
(e.g., West et al.\ 1995;
Allen et al.\ 1995;
West, Jones, \& Forman 1995).

\subsection{Radio Lobe Inflation or a Rotating Disk?} 
\label{sec:central_disk}

The central X-ray structure in the cD galaxy PKS2354-35
is similar to that in NGC~1275 (B\"{o}hringer et al.\ 1993).
In the case of NGC~1275, the regions of X-ray minima 
are filled by the radio lobes from the central cD galaxies. 
The radio nonthermal pressure in the lobes is comparable to
the X-ray thermal pressure in the surrounding gas, 
within the errors.
This is consistent with the idea that the radio lobes have displaced
the X-ray emitting gas, and are expanding against and being confined
by this gas.

The same model may apply to PKS2354-35 in A4059, but the radio emission
does not seem to extend completely into the X-ray holes
in either the 1.4 or 4.85 GHz radio images.
It is possible that the extended radio emission has
a very steep spectral index, and that lower frequency 
radio emission fills the X-ray holes.
Alternatively, it may be that the present period of radio activity in
PKS2354-35 is ending, 
or that the X-ray holes are relics of a previous
epoch of activity.
This might account for the absence of radio emission from the outer
parts of the X-ray holes.
In either scenario, the holes may still contain cosmic ray ions, low energy
cosmic ray electrons, and magnetic fields with a sufficient pressure
to maintain the X-ray holes against the pressure of the surrounding
X-ray emitting gas.

An alternative interpretation for the central X-ray structure emphasizes
the elongated X-ray bar rather than the X-ray holes.
Perhaps it is not that the X-ray emission from the holes is weak, but
rather that the emission orthogonal to the direction of the holes
(along the bar) is strong.
The X-ray bar might be a disk of X-ray emitting gas seen edge-on.
A disk might be formed from cooling flow gas with angular momentum, which
becomes more rotationally supported as it contacts into the center of the
cluster.
Eventually, part of the gas from this disk might feed an accretion disk
around the central AGN, and the radio jets may emerge perpendicular to
this disk.
Recently, 2-D numerical simulations have shown that
a rotationally supported disk 
can be formed under some circumstances in a cooling flow 
(Garasi, Loken, \& Burns 1996).
On the other hand, Nulsen, Stewart, \& Fabian (1984) have argued that
turbulent viscosity will effectively transport
the angular momentum of cooling flow gas outward, and that the gas will
not form a rotationally supported disk in the inner regions.
A disk might also be formed due to a merge with a gas rich
dwarf galaxy, although such galaxies are rare in centers
of rich cluster.

In either case, there would be no reason to expect the initial angular
momentum of the gas would be aligned with the major axis of the cluster
and cD galaxy.
However, West (1994) argues that cD galaxies are highly prolate, and
that differential procession will cause accreting gas to settle into
a disk which is perpendicular to the major axis of the galaxy, as
required in A4059.
Hydra A seems to show just such a disk, which is seen prominently
in 21 cm H~I emission
(Dwarakanat et al.\ 1995).
Optical emission line spectra of the disk in Hydra A show that it is
rotating 
(Ekers \& Simkin 1983;
Baum et al.\ 1988).
However, there is no similar evidence for a rotating disk
in Abell~4059.

\section{SUMMARY} \label{sec:summary}

In summary, we have analyzed the ROSAT HRI image and PSPC spectra
of Abell~4059.
The most striking feature of the X-ray image is
the two X-ray holes and an orthogonal X-ray bar
in the center of the cluster.
It is likely that the radio plasma from the cD galaxy
may have displaced the X-ray gas and created the X-ray holes,
or we may be seeing a rotationally supported disk of the 
X-ray emitting gas.
There may also be a moderate cooling flow 
in the center of the cluster,
$\dot{M} = 184^{+22}_{-25}$ $M_{\odot}$ yr$^{-1}$,
as suggested by the X-ray surface brightness profile.
However, such a cooling flow is not needed in
the spectral fitting of the PSPC data.
It may be that the cooling flow has just begun in the
center of the cluster.
In addition, very little intrinsic absorption was found 
in the center of the cluster.

\acknowledgments

This research was supported in part by NASA Astrophysical Theory Program
grant NAG 5-3057, NASA ASCA grant NAG 5-2526, and NASA ROSAT grant 5-3308.
The optical images are from the Digitized Sky Survey.
The Palomar Observatory Sky Survey was funded by the National Geographic
Society, and the Oschin Schmidt Telescope
operated by the California Institute of Technology and Palomar Observatory.
The Digitized Sky Survey was produced at the Space Telescope Science 
Institute (STScI) under U. S. Government grant NAG W-2166.

\end{document}